\documentclass{appolb}
\usepackage{epsfig}
\usepackage{cite}
\newcommand{\plaata}[4]{\raisebox{#2pt}{
\epsfig{figure=./#1.eps, width=#3cm,height=#4cm}}}
\newcommand{\psb}{\bar{\psi}}
\begin{document}

%\eqsec  % uncomment this line to get equations numbered by (sec.num)
\title{{\bf MONTE CARLO GENERATORS FOR THE LHC}
\thanks{Presented at the XXXI International Conference of Theoretical Physics,
Matter To The Deepest: Recent Developments In Physics of Fundamental 
Interactions, Ustro\'n, 5-11 September 2007, Poland.}
\thanks{Report number: \texttt{KA-TP-23-2007}} }
\author{Ma\l gorzata Worek 
\address{Institute for Theoretical Physics (ITP) \\Karlsruhe University, 76-128
  Karlsruhe, Germany\\
Institute of Physics, University of Silesia \\ Uniwersytecka 4, 40-007 
Katowice, Poland\\
\vspace{0.2cm}
e-mail: \texttt{worek@particle.uni-karlsruhe.de}
}}
\maketitle

\begin{abstract} 

The status of two Monte Carlo generators, \textsc{Helac-Phegas}, a program for 
multi-jet processes and \textsc{Vbfnlo}, a parton level program for vector boson fusion
processes at NLO QCD, is briefly presented. The aim  of these tools is the
simulation of events within the Standard Model at  current and future high
energy experiments, in particular the LHC.  Some results related to the
production of multi-jet final states at the LHC  are also shown.

\end{abstract}

The main aim of the Large Hadron Collider (LHC), which is expected to  start
in 2008, is the discovery of the last missing particle predicted by the
Standard Model (SM), the Higgs boson. Almost as high on the agenda, however,
is the search for  signals of new  physics beyond the SM. Background processes
to these searches are mostly  due to QCD interactions which are sometimes
accompanied by electroweak vector bosons. The final states are characterised
by a high number of jets and/or  identified particles. Theoretical predictions
in such cases require the computation of scattering amplitudes with a large
number of external particles.  The complexity of calculations grows with  the
number of external legs. For example, the numbers of  Feynman diagrams which
are needed for the computation of the  $gg \rightarrow 8g$ and $q\bar{q}
\rightarrow 8g$ amplitudes, are $10,525,900$ and $4,016,775$ respectively. In
general the number of Feynman diagrams grows asymptotically factorially with
the number of particles.  Moreover, for a given jet configuration there are
usually very many contributing subprocesses, e.g.  for the calculation of $pp
\rightarrow e^{+}\nu_{e}+ 6jets$, $2476$ subprocesses have to be taken into
account.  In addition neither the colour nor the spin of the partons are
observed.  Thus, for an amplitude  with p quarks and q gluons $(2\times
3)^{p}(2\times 8)^{q}$ configurations have  to be considered in principle for
every phase space point.  Both, the usual techniques of evaluating Feynman
diagrams and straightforward  summation over colour and helicity
configurations are in practice almost unusable. The next challenge  is the
phase-space integration. Each amplitude peaks in a complicated way inside the
momentum phase space. Direct integration is therefore impractical and one has
to search for efficient mappings to do importance sampling in a multi-particle
phase space. Clearly, new alternative techniques and automatisation of
calculations  for multileg LHC processes  is a timely task.

Over the last years new algorithms along with their implementations for
computing tree-order scattering amplitudes have been
proposed\cite{Caravaglios:1995cd,Draggiotis:1998gr,
Caravaglios:1998yr,Kanaki:2000ey,Draggiotis:2002hm,
Papadopoulos:2005ky,Cafarella:2007pc}.  They reorganise various off-shell
subamplitudes in a systematic way so that as little of the computation is
repeated as possible.  A scattering amplitude is computed through a set of
recursive equations derived from the effective action as a function of the
classical fields. These equations represent nothing else but the tree order
Dyson-Schwinger (DS) equations and give recursively the $n-$point Green's
functions in terms of the $1-$, $2-$,$\ldots$, $(n-1)-$point functions. They
hold all the information about the fields and their interactions for any
number of external legs and to all orders in perturbation theory.  For example
in QED these equations can be written as follows:
\[
\plaata{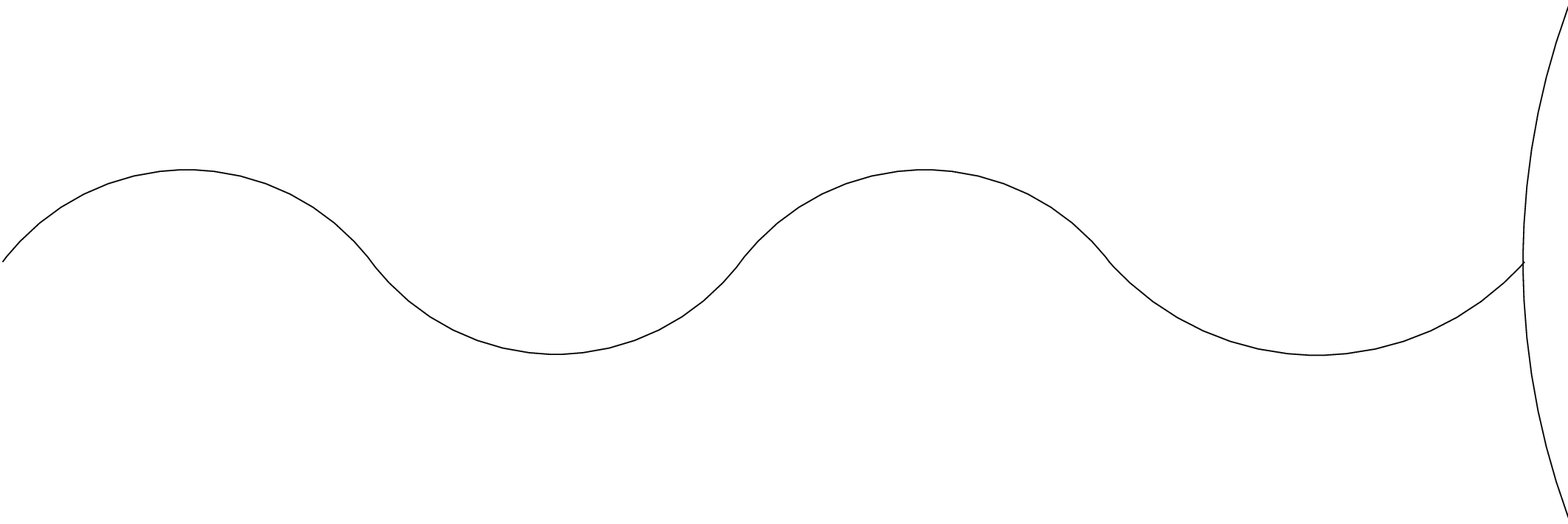}{-15}{1.8}{1}\;\;=\;\plaata{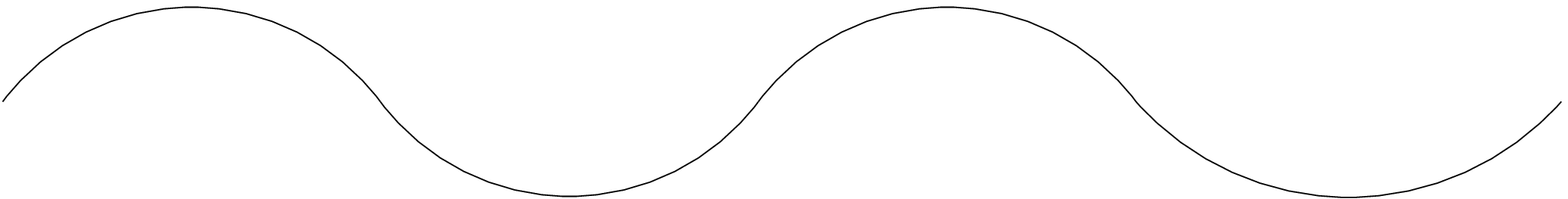}{0}{1.8}{0.2}\;
+\;\plaata{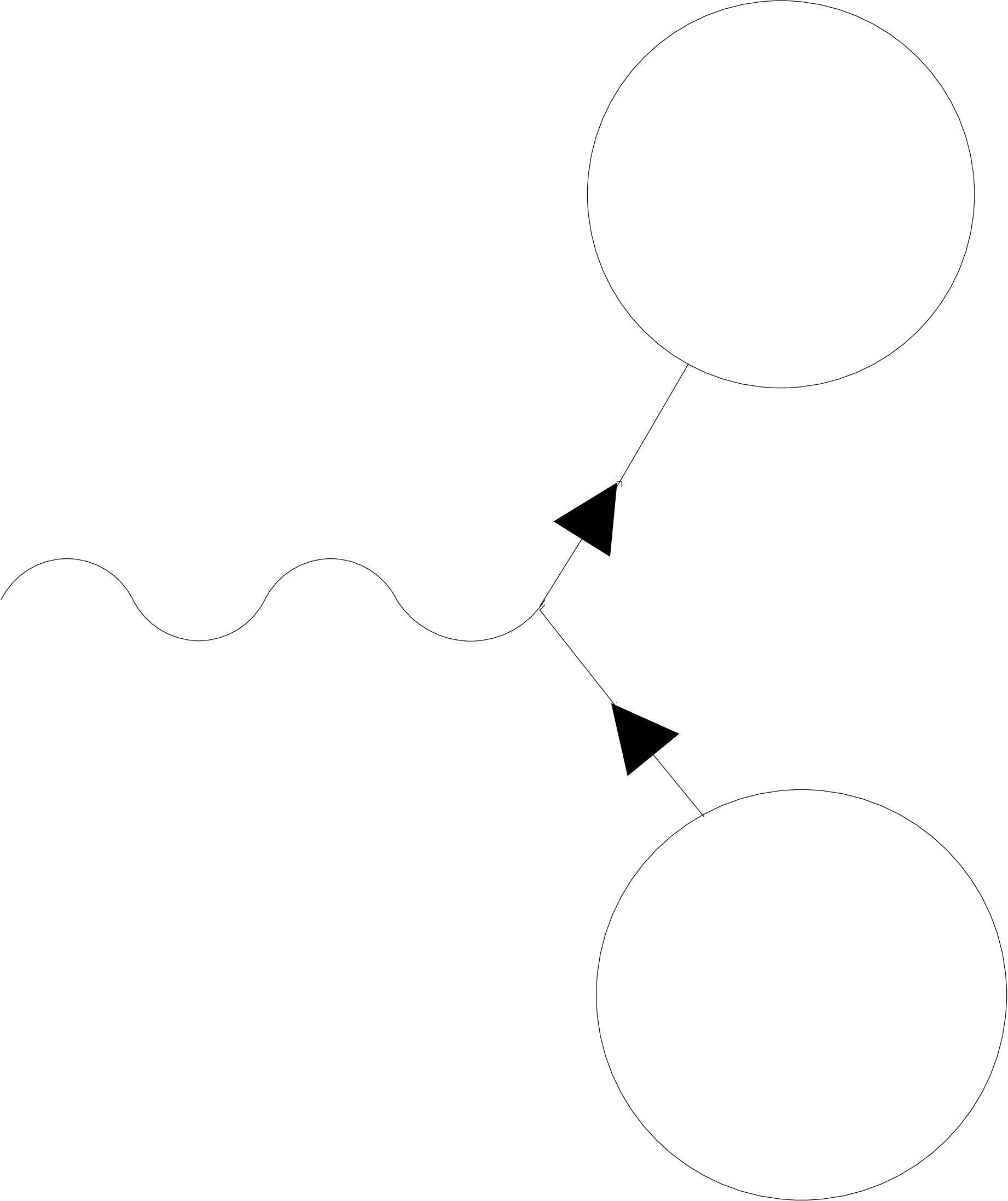}{-25}{1.8}{2}
\]
\begin{equation}
 b^\mu(P)=\sum_{i=1}^n \delta_{P=p_i} b^\mu(p_i) 
\sum_{P=P_1+P_2} (ig)\Pi^\mu_\nu (P_2)\gamma^\nu\psi(P_1)
\epsilon(P_1,P_2)
\end{equation}
where
\[
 b_\mu(P) \;=\; \plaata{bos1}{-5}{0.8}{0.5} \;\;\;\;
   \psi(P)  \;=\; \plaata{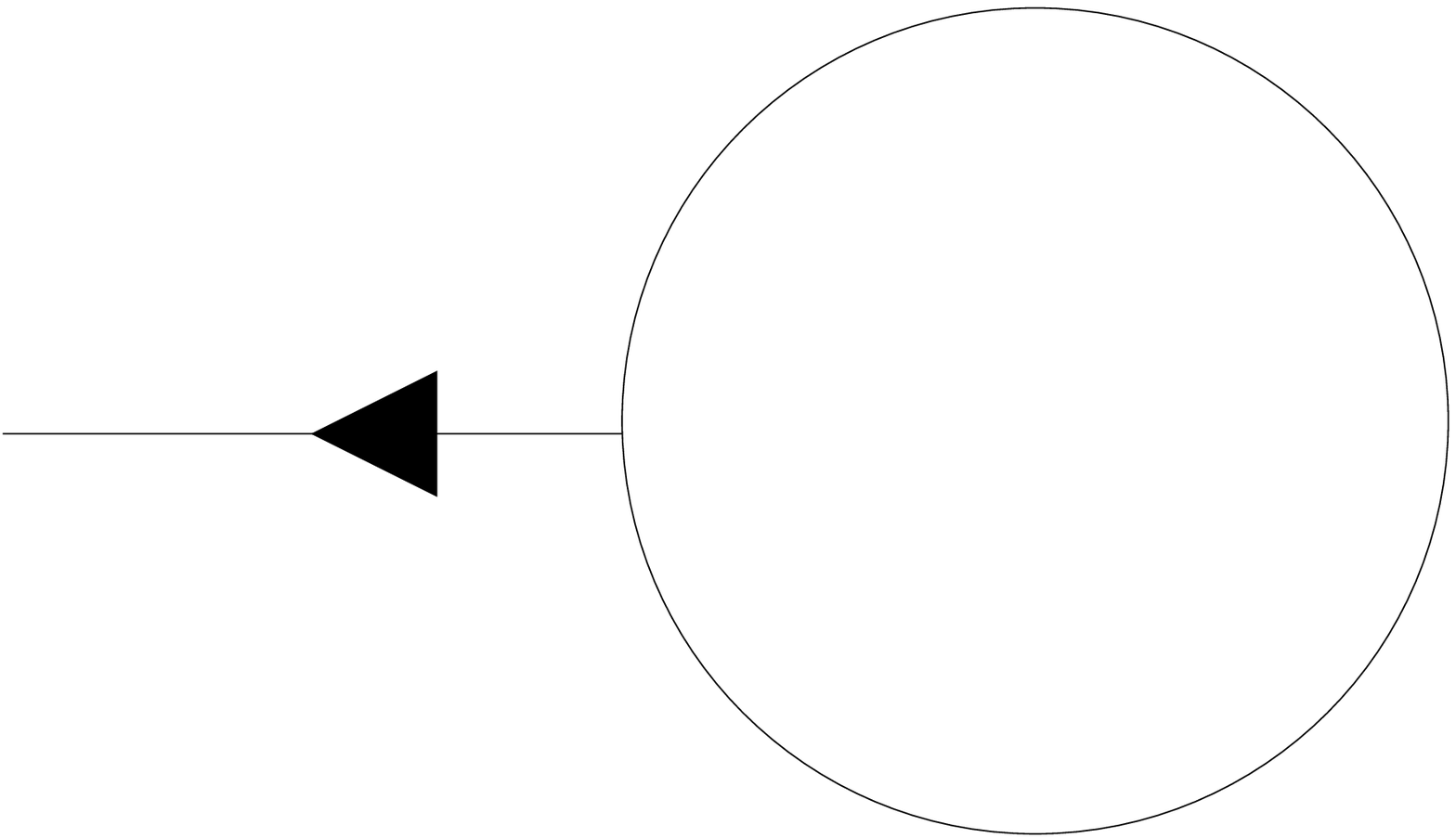}{-6}{0.8}{0.6}\; \;\;\;
   \psb(P)  \;=\; \plaata{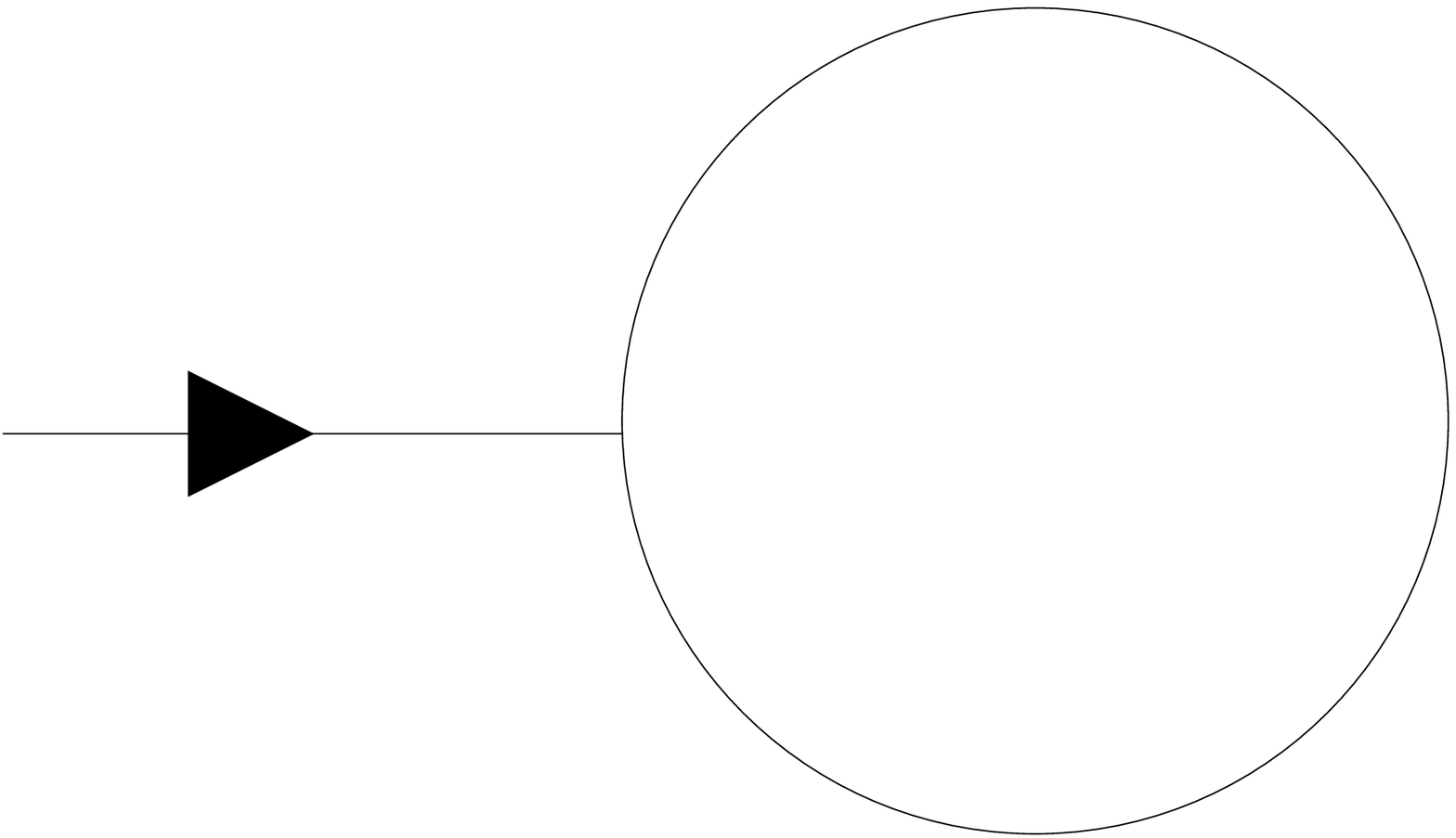}{-6}{0.8}{0.6}
\]    
describes a generic $n$-point Green's function with respectively one
outgoing photon, fermion or antifermion leg carrying momentum $P$.
$\Pi_{\mu\nu}$ stands for the boson propagator and $\epsilon$ takes into
account the sign due to fermion antisymmetrization. In the same way recursive
equations for other particles in  the SM  can be derived.
%*******************************************************************
\begin{figure}
\begin{center}  
\epsfig{file=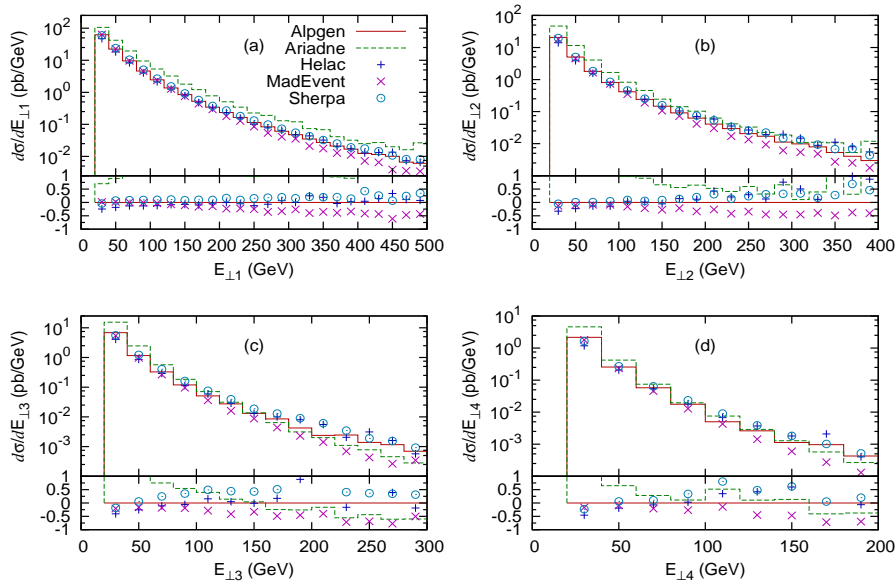,width=120mm,height=80mm}
\end{center} 
\caption  
{\it Inclusive $E_\perp$  spectra of the leading 4
  jets at the LHC (pb/GeV). In all cases the full line gives the
  \textsc{Alpgen} results, the dashed line gives the \textsc{Ariadne}
  results and the ``+'', ``x'' and ``o'' points give the \textsc{Helac},
  \textsc{Madevent} and \textsc{Sherpa} results respectively.}
\label{ptlhc}
\end{figure}  
%*******************************************************************
\begin{figure}
\begin{center}  
\epsfig{file=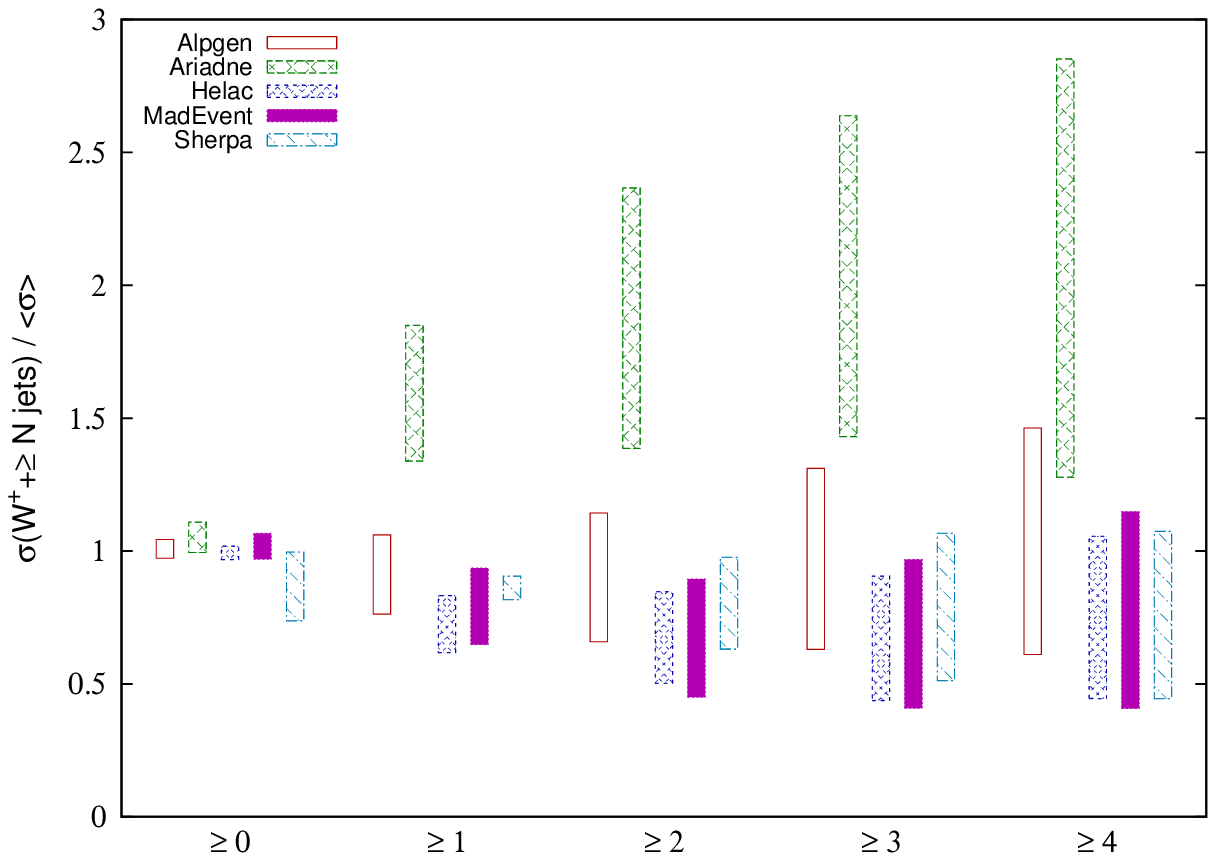,width=100mm,height=60mm}
\end{center} 
\caption  
{\it Range of variation for the LHC
  cross-section rates of the five codes, normalized to the average
  value of the default settings for all codes in each multiplicity
  bin.}
\label{rates}
\end{figure}  
%*******************************************************************

\textsc{Helac} \cite{Cafarella:2007pc}  is  the only existing implementation
of the algorithm based on DS equations. It is able to  calculate iteratively
matrix elements for an arbitrary multi-particle and multi-jet process within
the SM in leptonic and hadronic  collisions. For multi-jet states all
elementary parton level subprocesses are taken into account. All electroweak
vertices in both Feynman and unitary gauges have been included, whereas
unstable particles  are treated in a fully consistent way, by using either a
fixed width or  a complex mass scheme
\cite{Argyres:1995ym,Beenakker:1996kn,Denner:2006ic}. Spin and color
correlations are taken into account naturally and there is no approximation
involved. A substantial  speed up has been obtained with Monte Carlo (MC)
techniques to perform the sum over   helicity and color configurations
\cite{Draggiotis:2002hm,Papadopoulos:2005ky}.  The computational cost of
\textsc{Helac} grows like $\sim 4^n ~(3^n)$, which essentially counts the
steps used to solve the recursive equations\footnote{To reduce the
computational complexity down to an  asymptotic $3^n$ each 4-boson vertex must
be replaced with a 3-boson vertex \eg by introducing an auxiliary field
represented by the antisymmetric tensor $H^{\mu\nu}$, see
\cite{Draggiotis:2002hm,Papadopoulos:2005ky} for details.}.  The program
incorporates  the possibility to use  extended numerical precision by
exploiting the virtues of \textsc{Fortran90}. The user can easily switch to
quadruple precision or to an even higher, user-defined precision by using the
multi-precision library \cite{Smith:1991:AFP}. Finally, the peaking structure
of the amplitude is dealt with by the phase space generating algorithm
\textsc{Phegas}  \cite{Papadopoulos:2000tt}. \textsc{Phegas} is the first
implementation  of a completely automated algorithm of multi-channel phase
space mappings for  an arbitrary number of external particles. It uses the
information generated by \textsc{Helac} and automatically performs a
multi-channel phase space generation, utilising 'scalarized' Feynman
graphs. In the case of $pp$ and $p\bar{p}$  collisions the cross section is
also convoluted with parton distribution functions. In that case the
integration is  optimized by using the \textsc{Parni}  algorithm
\cite{vanHameren:2007pt}.   The program makes use of the Les Houches Accord
PDF Interface library (LHAPDF) \cite{Whalley:2005nh}. It also generates a Les
Houches Accord (LHA) file \cite{Boos:2001cv,Alwall:2006yp} with all the
necessary information needed to interface to the \textsc{Pythia}
\cite{Sjostrand:2006za} parton shower and hadronisation program. In fact, the
problem of  double counting of jets may arise when interfacing fixed order
tree level matrix elements to parton showers.  In order to deal with it, a
matching  algorithm has to be applied, which provides a smooth transition
between the part of the  phase space covered by parton showers and the one
described by matrix elements. We have used the so-called MLM matching
algorithm, see \eg \cite{Mangano:2006rw}. Let us note that  a comparative
study  \cite{Alwall:2007fs} of matching algorithms implemented in different MC
codes namely \textsc{Helac}, \textsc{Alpgen} \cite{Mangano:2002ea},
\textsc{Ariadne} \cite{Lonnblad:1992tz}, \textsc{MadEvent}
\cite{Maltoni:2002qb,Alwall:2007st} and \textsc{Sherpa}
\cite{Krauss:2001iv,Gleisberg:2003xi}  has recently been published for the
$W+n$ jets  production  with kinematics corresponding to the TeVatron and the
LHC.  As an example in Fig.\ref{ptlhc}, inclusive $E_\perp$ spectra of the
leading 4 jets at the LHC (pb/GeV) for \textsc{Alpgen}, \textsc{Ariadne}
\textsc{Helac}, \textsc{Madevent} and \textsc{Sherpa} are given.
Fig.\ref{rates} shows graphically the cross-section systematic error
ranges. For each multiplicity,  the rates are normalized to the average of the
default values of all the codes. The complete information on the simulation
details can be found in Ref.\cite{Alwall:2007fs}. The  \textsc{Helac-Phegas}
package \cite{Cafarella:2007pc} is now publicly available\footnote{
\texttt{http://helac-phegas.web.cern.ch/helac-phegas/}} see also
\cite{Kanaki:2000ms,Papadopoulos:2005vg,Papadopoulos:2005jv,
Papadopoulos:2006mh,Draggiotis:2006er}

As we have seen, if one is content with the tree level calculations  only,
it is possible to go to high orders with up to 8-10 partons in the final
state. Of course, they have to be kept  well separated to avoid the phase
space regions where divergencies become troublesome. Soft and collinear
regions can then be  covered by the parton shower. However, to resolve the
large scale dependence inherent in leading order calculations it is necessary
to include NLO corrections.  The complexity of a calculation increases with
the order in perturbation  theory.  Currently available NLO calculations are
restricted to 2-4 final state particles  only\footnote{There are no NLO
programs for the LHC with more than 3 hard particles in the final state. NLO
programs with four particles in the final state are available only for
$e^{+}e^{-}$ annihilation. See \eg \cite{Weinzierl:2007vk} for a recent review
on this subject.}.  More importantly, only one MC library, MC@NLO
\cite{Frixione:2006gn}, incorporates NLO QCD  matrix  elements consistently
into a parton shower framework.  A general purpose NLO MC library does not
exist yet. However, there are a few MC programs for specialised processes.

In particular, \textsc{Vbfnlo} belongs to this  category\footnote{
\texttt{http://www-itp.particle.uni-karlsruhe.de/$^{\sim}$vbfnloweb/}} when
various Vector Boson Fusion  (VBF) processes are concerned.
%*******************************************************************
\begin{figure}
\begin{center}  
\epsfig{file=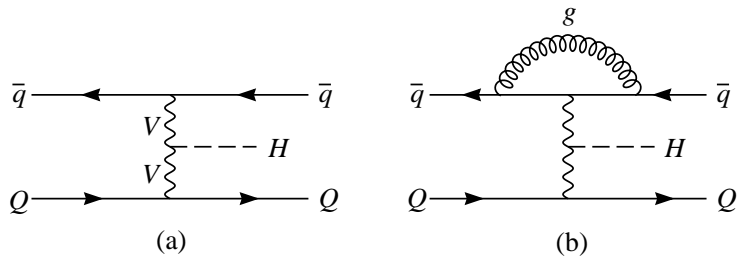,width=0.8\textwidth}
\end{center} 
\caption  
{\it Feynman graphs contributing to $\bar{q}Q\to \bar{q}Q H$ at
(a) tree level and (b) including virtual corrections to the upper quark line. }
\label{feynmandiagram}
\end{figure}
%*******************************************************************  For
For example, the $\bar{q}Q \rightarrow \bar{q} QH$ VBF process can be visualized
as the elastic scattering of two quarks mediated by the t-channel W or Z
exchange with the  Higgs boson radiated off the weak boson propagator,  see
Fig.\ref{feynmandiagram}. It is expected to provide a copious source of Higgs
bosons in pp collisions at the LHC and together with gluon fusion,  it
represents the most promising production process for Higgs boson
discovery. Once the Higgs boson has been found and its mass determined, the
measurement of its couplings to gauge bosons and fermions will be of main
interest. Here VBF will be of central importance since it allows for
independent observation in the  $H \rightarrow \gamma \gamma$, $H\rightarrow b
\bar{b}$, $H \rightarrow \tau^{+} \tau^{-}$, $H\rightarrow W^{+}W^{-}$ and  $H
\rightarrow invisible$ channels. This multitude of  channels is crucial for
separating the effects of different Higgs boson couplings. VBF measurements
can be performed at the LHC with statistical accuracies on cross sections
times decay branching ratios, $\sigma \times B$ reaching (5-10)\%
\cite{Zeppenfeld:2000td,Zeppenfeld:2002ng}.   Theoretical predictions of the
SM production  cross section with error  well below 10\% are required. This
clearly entails knowledge of the NLO QCD corrections.  In order to distinguish
the VBF Higgs boson signal from backgrounds, stringent cuts are required on
the Higgs boson decay products  as well as on the two forward quark jets which
are characteristic for VBF.  This can be best addressed with \textsc{Vbfnlo}
which  contains among others Higgs boson production  in the narrow resonance
approximation \cite{Figy:2003nv}.  In addition,  anomalous couplings have been
added for the Higgs boson \cite{Hankele:2006ma}.  The production of
$W\rightarrow l \nu_l $ and $Z \rightarrow l^{+}l^{-}$  \cite{Oleari:2003tc}
bosons in association with two jets is also included in the program since it
is an important background. Moreover, $W^+W^-$ \cite{Jager:2006zc} and $ZZ$
\cite{Jager:2006cp} production via vector-boson fusion  with subsequent
leptonic decay of the Ws and Zs with all resonant and non-resonant  Feynman
diagrams and spin correlations of the final-state leptons have been
implemented. Let us note that in all these  cases any identical fermion
effects, \ie $s$-channel exchange and interference  effects of $t$-channel and
$u$-channel diagrams are  systematically neglected. In the phase space  region
where VBF can be observed experimentally, with widely-separated quark  jets of
very large invariant mass, the neglected terms are strongly  suppressed by the
large momentum transfer in one or more  weak-boson propagators.  For the
evaluation of partonic matrix elements, amplitude techniques of
\cite{Hagiwara:1985yu,Hagiwara:1988pp} have been employed.  The calculation of
NLO QCD corrections is based on the  dipole subtraction formalism, in the
version proposed by Catani and Seymour \cite{Catani:1996vz}.  Radiative
corrections to a single quark line have only been calculated, since any
interference between subamplitudes with gluons attached to both the upper and
the lower quark lines vanishes identically at order $\alpha_s$,  because of
the color singlet nature of the exchanged weak boson. The virtual
contributions, obtained from the interference of one-loop diagrams with the
Born amplitude, include self-energy, triangle, box and pentagon corrections. A
Passarino-Veltman reduction of  tensor integrals \cite{Passarino:1978jh},
which is stable in the phase space regions covered by VBF-type reactions is
implemented up to box-type virtual corrections. For pentagon contributions,
however, this technique  gives rise to numerical instabilities, if kinematical
invariants, such as the Gram determinants, become small. Therefore the
reduction scheme proposed by Denner and Dittmaier for the tensor reduction of
pentagon integrals~\cite{Denner:2002ii,Denner:2005nn} has been used.  In all
cases the QCD corrections are modest,  changing total  cross sections by less
than 10\%. Remaining scale uncertainties  are at the few percent level. Modest
corrections are also present in distributions. Let us note that
\textsc{Vbfnlo} is a fully flexible MC program. Arbitrary cuts can be
implemented  and independent scales can be fixed for the  radiative correction
on the upper and lower quark lines.  Moreover, various scale choices and PDF
sets are available in the later case  also through the   LHAPDF library.
Finally, the program  generates an LHA file.

\vspace{1cm}
Work supported in part by the RTN European Programme MRTN-CT-2006-035505
HEPTOOLS - Tools and Precision Calculations for physics Discoveries  at
Colliders as well as by BMBF grant  05 HT6VKC. 
We would also like to thank the Galileo Galilei Institute  for
Theoretical Physics for the hospitality and the INFN for partial  support
during the completion of this work

%\bibliographystyle{h-elsevier3}
%\bibliography{ustron}

\end{document}